\documentclass[nohyperref]{article}

\usepackage{microtype}
\usepackage{graphicx}
\usepackage{subfigure}
\usepackage{booktabs} 
\usepackage{fancyvrb}

\usepackage{hyperref}

\usepackage[accepted]{icml2022}

\usepackage{amsmath}
\usepackage{amssymb}
\usepackage{mathtools}
\usepackage{amsthm}

\usepackage[capitalize,noabbrev]{cleveref}

\theoremstyle{plain}

\theoremstyle{definition}

\theoremstyle{remark}

\usepackage[textsize=tiny]{todonotes}

\icmltitlerunning{On pleasant semantics for differentiable programming languages}

\begin{document}

\twocolumn[
\icmltitle{Functional or imperative? On pleasant semantics for differentiable programming languages}




\begin{icmlauthorlist}
\icmlauthor{Michael Innes}{indy}
\end{icmlauthorlist}

\icmlaffiliation{indy}{Independent}

\icmlcorrespondingauthor{Michael Innes}{mike.j.innes@gmail.com}

\icmlkeywords{Machine Learning, ICML}

\vskip 0.3in
]



\printAffiliationsAndNotice{}  

\begin{abstract}
In machine learning (ML), researchers and engineers seem to be at odds. System implementers would prefer models to be declarative, with detailed type information and semantic restrictions that allow models to be optimised, rearranged and parallelised. Yet practitioners show an overwhelming preference for dynamic, imperative languages with mutable state, and much engineering effort is spent bridging the resulting semantic divide. Is there a fundamental conflict? This article explores why imperative and functional styles are used, and how future language designs might get the best of both worlds.
\end{abstract}

\section{Why functional?}

General purpose, numerical programming languages are the right tools for building models in machine learning, probabilistic programming, and differentiable programming ($\partial P$). In general, models may be no less complex than general-purpose programs, and need similar tools to manage that complexity \cite{innes2018machine}.

ML practitioners tend to use imperative languages -- mainly Python, and to a lesser extent R, Matlab, Julia and a smattering of others -- for their work. Yet ML, and particularly deep learning, frameworks often look quite unlike normal libraries, imposing an atypical functional style of programming \cite{colah2015types}.

The term ``functional'' originated with the use of higher-order functions, and is now associated with a number of related ideas. Two that matter for our purposes are:

\begin{enumerate}
\item \textit{Value semantics}, particularly for arrays. Mathematical operations create new arrays, rather than operating in-place.
\item \textit{Referential transparency}, in which a function call can be replaced by its return value. In other words, implicit dependencies (such as model configuration objects) can be assumed fixed.
\end{enumerate}

JAX, a Python library for numerics and code transforms \cite{frostig2018compiling}, is a clean example of these issues at play. It turns the Python interpreter into a compiler, using objects that abstractly represent arrays to run a partial evaluation of user code. In effect the result of dictionary lookups and dynamic dispatch is cached on first run, which avoids overhead during model training. But strictly speaking this is in violation of Python's semantics: the result of a dictionary lookup, or almost any other Python operation, might well change at run time. So JAX simply asks users to provide a pure function and assumes that they have done so, with undefined behaviour if the rule is violated. This has some limitations and sharp edges that would be ameliorated with better language-level support.

Julia, a numerical programming language \cite{bezanson2017julia}, is revealing in a different way. Julia's abstract interpretation is analogous to JAX's partial evaluation (a proof-of-concept hybrid, \citealt{innes2020mjolnir}, showed that they are two sides of the same coin). But Julia's compiler must respect the language's semantics, without making additional assumptions. Mutable objects, such as \texttt{Dict}s, stymie inference and must be given explicit type parameters to allow inference of downstream code. Limited constant propagation is supported on immutable objects, like \texttt{Tuple}s, but not arrays or dictionaries. Julia's source-transform based autodiff library, Zygote, disables array mutation in user functions due to the complexity of managing their derivatives. Deep learning with Flux, a Julia-based ML stack \cite{innes2018fashionable}, therefore looks functional: but in this case the core language's use of mutation impedes, rather than enabling, high performance code.

The trend towards functional-style, relatively declarative programming with compilers like XLA \cite{sabne2020xla} is a significant shift from the conventional style of scientific and numerical computing,\footnote{And indeed the style Julia was designed to cater for, albeit with more modern language affordances.} in which arrays are explicitly allocated and modified in loops, with a close correspondence between program and machine instructions. The change is partly driven by increasingly diverse hardware, as well as the relatively fixed set of core numerical kernels used in ML. Kernels individually benefit from hand optimisation, but compositions of them have less room for improvement, even for hardware experts. Low-level control is therefore less valuable for users, and general optimisations like simple loop fusion, memory placement and model-level distribution and parallelism are best done by compilers, on representations far more abstract than machine code \cite{li2020deep}.

\section{Why imperative?}

The discussion above seems to suggest that Haskell, OCaml or Scala would be better frontends for numerical compilers. Yet researchers have voted with their feet (or at least with their keyboards): Python is by far the most popular language for ML,\footnote{Google Trends supports ``python machine learning'' being about 10$\times$ more popular than ``matlab machine learning'', which in turn is about 5$\times$ more searched than ``julia machine learning''.} despite being poorly suited to the task on paper. Why are dynamic, imperative languages so popular? Partly this is to do with conventional (though not inherent) differences between static and dynamic languages -- things like interactivity, ease of use for beginners and the flexibility of dictionary-oriented programming \cite{dubois1996numerical}. We don't delve into those topics, except to point out that there are perfectly serviceable dynamic, functional languages like Clojure and Mathematica, so it can't be a complete explanation.\footnote{There are also some historical accidents that favour Python specifically. For example, its unusual choice to use reference counting happens to be ideal for managing large GPU arrays, where Julia's tracing GC struggles.}

Given that ML frameworks eschew shared mutable state, the last difference of note between functional and imperative is their conventional syntax choices. Clearly, nested, structured control flow (as in the loop, \texttt{if}/\texttt{else} statements and \texttt{break}, \texttt{return} and \texttt{continue}) is a helpful way to express algorithms, especially the iterative algorithms common in numerical programming. Just as importantly, the imperative style allows for a notion of \textit{change} that is easier to think with, compared with having to decompose and reconstruct everything: ``change column $i$ of matrix $A$'' is a more natural description than ``take all columns but the $i^{th}$ from $A$, and concatenate them together with this new one in the right position to create $B$'', and more so as updates get more complex.

Syntax like \texttt{A[i] = x} typically does three things: it provides a syntactic notion of change as above; it efficiently modifies memory in-place; and it may alter the behaviour of seventeen other threads of execution that reference $A$. But the last of these is most often an anti-feature, and the second is the job of the compiler. It really is only the syntax that we want.

Despite being superficial in a sense, syntax is an important part of the user experience, and a perfectly good reason for users to prefer imperative over functional languages. But now we are surely asking to have our cake and eat it too: we don't like mutable state, yet we want to write programs with explicit change. In fact, there is no contradiction. To explain why we'll have to challenge some misconceptions about ``imperative'' syntax.

\section{The procedural is pure}

When discussing imperative programming, it's common to conflate mutable \textit{local variables} with mutable \textit{data structures}. In fact they are hardly related. Here's an example with the former, but not the latter:

\vspace{-0.16cm}
\begin{Verbatim}[fontsize=\footnotesize]
fn pow(x, n) {
  r = 1
  while n > 0 {
    r *= x
    n -= 1
  }
  return r
}
\end{Verbatim}
\vspace{-0.16cm}

LLVM, a compiler for those most imperative of languages \cite{lattner2004llvm}, lowers such functions to an intermediate representation (IR) similar to the following:

\vspace{-0.16cm}
\begin{Verbatim}[fontsize=\footnotesize]
fn pow(x, n) {
  fn block1() {
    r = 1
    block2(n, r)
  }
  fn block2(n, r) {
    if n > 0 {
      block3(n, r)
    else {
      block4(r)
    }
  }
  fn block3(n, r) {
    n_ = n - 1
    r_ = r * x
    block2(n_, r_)
  }
  fn block4(r) {
    return r
  }
  block1()
}
\end{Verbatim}
\vspace{-0.16cm}

LLVM and similar compilers call this form ``static single assignment'', or SSA, because variables are syntactically immutable -- in other words, they behave like bindings in functional languages. We could equally call this a set of mutually tail-recursive pure functions \cite{appel1998ssa}. This form is excellent for a wide variety of analysis, optimisation and transformation (eg autodiff on SSA, \citealt{innes2020sense}). All local control flow, however complex, can be lowered to SSA \cite{cytron1989efficient}.

The main point is that ``mutable'' locals are a corollary of control flow structures, rather than of other kinds of mutation. And control flow is just syntax sugar for simple functions. If syntax sugar does not fundamentally change our programming paradigm, and the latter example is a functional program, the former must be too. At the least, we can agree that mutable locals need not violate referential transparency, because there's no external way to tell they are being used.\footnote{Shared mutable state can also be emulated in a functional language. But a more invasive, global transformation would be needed to capture all of its ``features'', like race conditions and nondeterminism, so it represents a more fundamental difference.}

It is right to say that variables like $r$ are changing. But we use the word \textit{change} in the sense of changing clothes, without assuming any physical alteration.\footnote{\textit{change} derives from words meaning ``barter'' or ``substitute''; ``cause to turn or pass from one state to another''.} A change might compile down to a real modification (of registers, memory or whatever) but this is only coincidence. Change and mutation are distinct; C just happens to conflate them.

\section{The more things change...}

The exploration above suggests that clearer notions of change may help us get the benefits of imperative-style programming, without losing those of the functional style. But simply adding control flow and locals to a functional language is not enough. For one thing, it is clunky to write

\vspace{-0.16cm}
\begin{Verbatim}[fontsize=\footnotesize]
xs = append(xs, x)
\end{Verbatim}
\vspace{-0.16cm}

every time we want to change $xs$ (in this case, by appending a new element to the list). It has the benefit of being explicit about what is changing, but at the cost of some verbosity, especially when multiple inputs are affected.

We propose a design that takes inspiration from the C-like address operator \texttt{\&}, writing this as

\vspace{-0.16cm}
\begin{Verbatim}[fontsize=\footnotesize]
append(&xs, x)
\end{Verbatim}
\vspace{-0.16cm}

which is still explicit, but less verbose. We refer to \texttt{\&} as the ``swap operator''. All function calls have a hidden return value, a list of swapped arguments, so that for example

\vspace{-0.16cm}
\begin{Verbatim}[fontsize=\footnotesize]
result = foo(&a, b, &c)
\end{Verbatim}
\vspace{-0.16cm}

is roughly equivalent to

\vspace{-0.16cm}
\begin{Verbatim}[fontsize=\footnotesize]
[result, rs] = foo(a, b, c)
a = rs[1]
c = rs[3]
\end{Verbatim}
\vspace{-0.16cm}

We can write a function to swap two variables as follows:

\vspace{-0.16cm}
\begin{Verbatim}[fontsize=\footnotesize]
fn switch(&x, &y) {
  [x, y] = [y, x]
  return
}

a = 1
b = 2
switch(&a, &b) # now a = 2, b = 1
\end{Verbatim}
\vspace{-0.16cm}

where the \texttt{\&a} in the function signature indicates that the final value of \texttt{a} should be returned to the caller.

The use of \texttt{\&} is like the address operator in C, C++, Rust and Swift \cite{ritchie1988c}. But whereas the \textit{change} implied is similar, those languages conflate that change with \textit{mutation} of the stack, with all the knock-on effects that implies. In our implementation no references, pointers or mutable state are created; the \texttt{\&} is syntax sugar for a functional program, which means that program transformations and analyses are not impeded.

A last sprinkle help us recover full convenience. Syntax like \texttt{A[i] = x} and \texttt{foo.bar = baz} can lower to function calls (eg \texttt{set(\&A, x, i})). We can update nested structures like \texttt{append(\&foo.xs, x)}, which roughly becomes:

\vspace{-0.16cm}
\begin{Verbatim}[fontsize=\footnotesize]
xs = foo.xs
append(&xs, x)
foo.xs = xs
\end{Verbatim}
\vspace{-0.16cm}

This looks imperative. But only variables are changing -- not values.

Note that in most languages with mutable locals, closures capture variables by reference. In this more functional setting they should of course capture by value instead. This is more intuitive in many cases anyway (eg a closure that captures a loop counter by reference can lead to confusing bugs).

A side effect of this approach is to avoid a leaky abstraction most imperative language have: types that happen to be implemented as heap-allocated references, like dictionaries or objects, are mutable, whereas basic types like numbers or strings are not. More generally, possible modifications are dictated by the underlying representation in memory, and the choice to make an interface in-place or not may be dictated by implementation concerns. When change and mutation are distinct, any value is just as changeable as any other.

With these relatively minor tweaks -- structured control flow, mutable locals and a bit of syntax sugar -- we can write numerical programs almost exactly as we would in Python or Julia. But because our data structures are values, we have the guarantees needed for aggressive compilation and transformation by default. Such optimisation can be done robustly by the main compiler, without special cases or sharp edges, and while playing nicely with existing language tooling: error messages, debuggers and profilers.

\section{Turning change into mutation}

One reason that change, in the sense we have discussed, and mutation are conflated is that doing things in-place is a good optimisation. Where performance matters, it is crucial that this optimisation be predictable to the programmer. If it isn't, algorithms may not only have unnecessary overheads but also higher computational complexity (because mutations are usually constant time, whereas copies are linear). Having to make copies is therefore a significant downside of the functional style.

It is not fatal, however. Reference counting (and its compile-time analogues, as in Rust or Lobster, \citealt{matsakis2014rust,lobster}) provides a solution: any time the count is $1$, no-one else can see our mutations, so it is safe to update the old data in place.

Koka \cite{leijen2016algebraic} has an implementation of this idea, and it turns out to benefit from the functional pattern of deconstructing and rebuilding data. For example, say we have a type \texttt{Foo} with named fields \texttt{xs}, \texttt{ys} and \texttt{zs}. We want to append a new item, \texttt{x}, to the \texttt{xs} array. In a functional style we might write the update something like this, using pattern matching to destructure \texttt{foo}:

\vspace{-0.16cm}
\begin{Verbatim}[fontsize=\footnotesize]
Foo(xs, ys, zs) = foo
xs2 = append(xs, x)
foo2 = Foo(xs2, ys, zs)
\end{Verbatim}
\vspace{-0.16cm}

After the first line has executed, \texttt{foo} is not used, and can therefore be released. Assuming \texttt{foo} is unique, releasing it will in turn release its hold on \texttt{xs}. \texttt{xs} is now only referenced from one place (the current stack frame) and can safely be updated by mutation.

Consider in contrast the version we wrote above, using an imperative-style update syntax to change \texttt{foo}:

\vspace{-0.16cm}
\begin{Verbatim}[fontsize=\footnotesize]
xs = foo.xs
append(&xs, x)
foo.xs = xs
\end{Verbatim}
\vspace{-0.16cm}

Unlike the last example, \texttt{foo} is live after the first line, because we need it to run the last line. That in turns means \texttt{xs} is referenced in two places (the current stack frame, and \texttt{foo}) and can't be modified in place.

Fortunately, the swap syntax we introduced for nested updates, \texttt{append(\&foo.xs, x)}, can also account for this case. Rather than expanding directly to the above, it expands to code that decomposes and rebuilds \texttt{foo} more like the functional-style example. This can work recursively, meaning that updates to deeply-nested structures will still reliably happen in place where possible.

\section{Related work}

There are many interesting efforts to bridge the gap between functional and imperative, including within numerical and differentiable programming.

Dex \cite{maclaurin2019dex} aims to do numerical programming and code transformation in a Haskell-like setting, notably using constructs for structured mutation inspired by algebraic effects.

Koka \cite{leijen2016algebraic} is a recent functional language with support for side effects based on algebraic effect handlers. Notably, effects imply that order matters, so Koka uses a C-like syntax for sequences of statements.

Clojure \cite{hickey2008clojure} is related for somewhat tangential reasons. A key benefit of languages like Python is the ease of ``dictionary-oriented programming'', where the built-in data types are flexible enough that users typically don't have to build their own. Clojure showed that this was practical in a functional setting with its persistent data structures, which will be an important component of any system resembling a ``functional Python''.

Rust \cite{matsakis2014rust} takes a different tack to eliminating shared mutable state, not by avoiding mutation but by restricting sharing through static analysis. This approach is probably inappropriate for our use case, but Rust libraries have thought carefully about how to express more advanced techniques, such as multiple threads working on an output array, in a structured and safe way. Those APIs could inspire functional approaches that can reliably compile to efficient code.

\section{Conclusion}

The functional and imperative styles are less at odds than they may first appear. At least in machine learning, the functional approach is valuable for its semantic benefits, while imperative style brings syntactic and expressive convenience. And though the benefits of FP are real, there is no need to make users do SSA transformation by hand. Perhaps surprisingly, it is possible to construct a very imperative \textit{looking} language that nonetheless retains the most important functional properties. Such designs are promising for probabilistic and differentiable programming, offering high flexibility to users expressing models, without sacrificing the ability to do advanced analysis, optimisation and transformation of the resulting code.

\textit{Thanks to Zenna Tavares for encouraging this paper, and to Maximilian Schleich for reviewing a draft.}


\bibliography{main}
\bibliographystyle{icml2022}



\end{document}